\documentclass[a4paper,12pt]{article}
\usepackage{geometry}
\usepackage[utf8]{inputenc}
\usepackage[T1]{fontenc}
\pdfoutput=1
\usepackage{amsmath}
\usepackage{amssymb}
\usepackage{amsfonts}
\usepackage{lmodern}
\usepackage{tikz}
\usepackage{pgfplots}
\usepackage{rotating}
\usepackage{xcolor}
\usepackage{graphicx}
\usepackage{tgheros}
\usepackage{helvet}
\usepackage{amssymb}
\usepackage{tikz}
\usepackage{csquotes}
\usepackage[english]{babel}
\usepackage[]{biblatex}
\usepackage{wrapfig}
\usepackage{float}
\usepackage{multicol}
\usepackage{subcaption}
\usepackage{booktabs}
\usepackage{setspace}
\usepackage{blindtext}
\usepackage[margin=1cm]{caption}
\usepackage{longtable}
\usepackage{layouts}
\definecolor{blue}{RGB}{ 25 , 25, 112}
\definecolor{red}{RGB}{255,0,0}
\definecolor{almond}{rgb}{0.94, 0.87, 0.8}

\usetikzlibrary{matrix,positioning,calc}
\usetikzlibrary{backgrounds}
\usetikzlibrary{intersections}
\usetikzlibrary{calc}
\usetikzlibrary{arrows}
\usetikzlibrary{backgrounds}
\usetikzlibrary{intersections}
\usetikzlibrary{matrix,positioning,calc}
\usetikzlibrary{backgrounds}
\usetikzlibrary{intersections}
\usetikzlibrary{calc}
\usetikzlibrary{arrows}
\usetikzlibrary{shadows}
\usetikzlibrary{backgrounds}
\usetikzlibrary{intersections}
\usetikzlibrary{arrows}

\title{Coupling the Gini and Angles to Evaluate Economic Dispersion}
\author{Mario Schlemmer}
\date{}
\begin{document}
	\maketitle
	\thispagestyle{empty}
\begin{flushleft}
	
	Classical measures of inequality use the mean as the benchmark of economic dispersion. They are not sensitive to inequality at the left tail of the distribution, where it would matter most. This paper presents a new inequality measurement tool that gives more weight to inequality at the lower end of the distribution, it is based on the comparison of all value pairs and synthesizes the dispersion of the whole distribution. The differences that sum to the Gini coefficient are scaled by angular differences between observations. The resulting index possesses a set of desirable properties, including normalization, scale invariance, population invariance, transfer sensitivity, and weak decomposability.
	
\end{flushleft} 
\textbf{JEL-classication:} D31, D63

\section*{Introduction}
In recent years, there has been increased interest in inequality measures that don't rely on the mean of the distribution. In general, the focus of classical measures like the Gini coefficient is on the concentration aspect of inequality, facilitated by increasingly skewed distributions this tendency has increased over the last decades. The focus on the right half makes these mean-based measures insensitive toward inequality at the lower end of the distribution.
To overcome these limitations different measures have been proposed that take the relative nature of the poor and the rich into account. Among them are inter-decile ratios like the Palma which contrasts the income of the richest 10 percent and the poorest 40 percent(Cobham and Sumner, 2013). But this measure does not take the whole distribution into account and such a grouping of values at the lower end of the distribution is not sensitive if the lowest deciles have far less then the 4th decile. Some measures that are members of the General Entropy class take the whole distribution into account and are more sensitive to inequality at the left tail, but they can not accommodate zero values or negative values. A new inequality measure that is sensitive to inequality at the lower end of the distribution, that can be used if there are zero values and negative values and that satisfies the main axioms of inequality measures can be obtained by combining differentials that sum to the classical Gini index with angular differences between observations.

The Gini coefficient is usually defined in terms of the Lorenz curve,  another popular formula often encountered in the literature is the mean absolute difference $\Delta$ divided by the mean of the distribution. The popular index can also be directly expressed as a sum of contributing differences between value pairs, for $n$ observations $Y_n = \{y_1, y_2, ..., y_n\}$ in distribution $F$, the Gini it is defined as:
\begin{align}
	\begin{split}
	\text{G}=\sum_{comb.}\text{D}\text{\small{, where}}
	\end{split}\\
	\begin{split}\nonumber
	\text{D}=\frac{|y_i-y_j|}{n^2\mu}
	\end{split}
\end{align}
Each \text{D} is identical to the contribution of a value pair to the Gini coefficient and therefore also corresponds to its share of the area between the Lorenz curve and the 45-degree line.  For the value of \text{D}, it does not matter if the difference is between two small or two large values, only the absolute difference relative to the mean counts. As a result of this, the Gini is relatively insensitive to very low values including zero and negative values. A proportional differential that takes the size of the compared values into account is given by: 
\begin{align}
	\begin{split}
		\angle =\frac{2}{\pi} \Big | atan2(y_i,y_j) - atan2(y_j,y_i)\Big| 
	\end{split}
\end{align}
 The value of the angular difference $\angle$ is bound between 0 and 1. It is a measure of proportional difference, the same absolute difference is evaluated differently depending on how large the compared values are. The average value of $\angle$ for all pairwise comparisons is a measure of dispersion that lacks the crucial property of transfer sensitivity. A progressive transfer from the top to the middle of the distribution can increase its value. But by combining the angular difference $\angle$ with D its sensitivity can be utilized to develop an inequality measure.The absolute difference between $atan2(y_i,y_j)$ and $atan2(y_j,y_i)$ is used in Eq.2 because the order of arguments $(y_i,y_j)$ for the function $atan2$ is not the same in every software, taking the absolute difference always yields the correct results.

\section*{Combining Gini's and proportional differences}
The new measure can be expressed as the product of D and $\angle$ for all combinations of observations in the distribution: 
\begin{align}	
	\text{V}= \sum_{comb.}\text{D} \cdot \angle
\end{align}	
 Index values obtained with V are lower than those obtained with the Gini. A pairwise comparison will only contribute a high value if both D and $\angle$ indicate a high level of inequality. Large proportional differences between two small values do not contribute much, but the same is true for large absolute differences between two values that lack a substantial proportional difference. This dual evaluation of differences emphasizes differences between the fortunate and the less fortunate in a distribution. Absolute differences between fortunate individuals, often of less social concern, have less influence. The measure is normalized, such that the index value lies between 0 and 1. If all incomes are the same \text{V}=0. For finite $n$ the upper bound is given by ${(n-1)}/{n}$. The proposed measure also satisfies the following normative axioms that have been proposed to aid in identifying appropriate inequality indices:\\
\\
\indent 1. \emph{Scale invariance:} The index value does not change if all incomes are changed\par proportionally.\\
\indent 2. \emph{Population invariance:} The index value does not change if the original popula-\par tion is replicated.\\
\indent 3. \emph{Pigou–Dalton principle of transfers:} If a progressive transfer does not change\par the rank of the individuals in the distribution the index value decreases.\\
\indent4. \emph{Principle of strong diminishing transfer:} This principle requires that if there are\par two income pairs with the same absolute income difference in the distribution, that\par the inequality reducing effect of a progressive transfer is stronger if it occurs be-\par tween the income pair that has a lower share of total income (Kolm,1976). \\

\indent V can accommodate zero values, in most programs the angular difference between two zero values is defined as zero. The measure can also be used if there are negative values, up to a certain degree. The Pigou-Dalton transfer will be satisfied as long as the proportion of negative and zero values in the distribution is less than half of all observations. Given that the proportion of such values is usually lower in empirical distributions of income and wealth this limitation should not pose a problem in most cases. As a benefit, common methods that are when inequality is calculated with some other measures like setting negative values to zero or adding 1 to all values are not needed if inequality is quantified with the proposed measure. Furthermore, V satisfies a weak form of subgroup decomposability(Ebert,2010). The index value for the total population can be expressed as the weighted sum of within-group terms that are based on index V for the subgroups ($\text{V}_{\text{G1}}$, $\text{V}_{\text{G2}}$), and a between-group term that is based on all comparisons of values between members of different subgroups. The calculation for large data sets is computationally expensive, but it is pretty straightforward for percentiles and other quantiles. The good econometric properties make the proposed measure a viable alternative to existing measures, it can be used in econometrics for the analysis of income and wealth inequality, but also in other disciplines.

\newpage
\section*{References}
\normalsize{Cobham, A., \& Sumner, A. (2013). Is it all about the tails? The Palma measure of\par income inequality. \textit{Center for Global Development working paper}, (343)}.\\
\normalsize{Ebert, U. (2010). The decomposition of inequality reconsidered: Weakly decomposable\par measures. \textit{Mathematical Social Sciences, 60}(2), 94-103.}\\
\normalsize{Kolm, S. C. (1976). Unequal inequalities. II. Journal of economic theory, 13(1), 82-111.}\\
\end{document}